\documentclass{article}

\usepackage{arxiv}

\usepackage[utf8]{inputenc} 
\usepackage[T1]{fontenc}    
\usepackage{hyperref}       
\usepackage{url}            
\usepackage{booktabs}       
\usepackage{amsfonts}       
\usepackage{nicefrac}       
\usepackage{microtype}      
\usepackage{lipsum}		
\usepackage{graphicx}
\usepackage{natbib}
\usepackage{doi}
\usepackage{setspace}

\title{Coloring Inside the Lines: The Jagged Legacy of the HOLC Neighborhood Risk Maps}

\date{March 9, 2022}	

\author{ \href{https://orcid.org/0000-0001-9073-1349}{\includegraphics[scale=0.06]{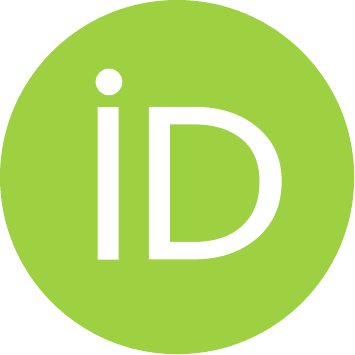}\hspace{1mm}Arunav Gupta}\thanks{The author would like to thank Dr. Isaac Martin for advising this work and HDSI for the scholarship that funded it.} \\
	Halıcıoğlu Data Science Institute\\
	University of California San Diego\\
	San Diego, CA\\
	\texttt{arg002@ucsd.edu}\\
}

\renewcommand{\shorttitle}{Coloring Inside the Lines}

\hypersetup{
pdftitle={Coloring Inside the Lines},
pdfsubject={stat.AP},
pdfauthor={Arunav Gupta},
pdfkeywords={redlining, HOLC, housing discrimination, markov chain monte carlo},
}

\begin{document}
\maketitle

\begin{abstract}
	There has been a large body of work exploring the discriminatory nature of the home mortgage risk maps produced by the Home Owners’ Loan Corporation in the late 1930s. However, little attention has been paid to the question of whether these maps are still descriptive of racial residential boundaries in their cities 80 years after their creation. To address this gap, Markov Chain Monte Carlo, previously unutilized in the relevant literature, is employed to randomly generate many plausible alternative mortgage security maps. Then, the racial evenness of the HOLC maps and the generated maps is compared using Shannon’s entropy. These findings indicate that the HOLC maps are significantly descriptive of the precise racial residential boundaries prevalent across eleven US cities in 2010. The methodology used here is highly modular and reproducible, allowing for future work measuring different outcome statistics, locations, and time periods.
\end{abstract}


\section{Introduction}

In the midst of the Great Depression in the 1930’s, President Roosevelt’s Administration was tasked with preventing the foreclosures of homes in American cities while many citizens struggled to recover from the economic impacts and job loss of the Depression. Roosevelt’s solution to the problem included the Home Owners’ Loan Corporation (HOLC), which, along with the Federal Housing Administration, was created in order to stave off declining demand in the housing market and an increase in overall lending risk caused by mass unemployment and the collapse of key financial institutions. These groups had two purposes: first, to insure loans made by lending institutions to shore up confidence in the lending market; second, to refinance existing home loans using a standardized, uniform method of risk appraisal (\citet{fishback10}). Both goals manifest themselves in the HOLC redlining maps, at the time known as neighborhood risk maps. 

There is a large body of work surrounding the causes and effects of redlining, more specifically the HOLC maps (\citet{Hillier2003}, \citet{Hillier2005}, \citet{Immergluck2002}, \citet{fishback14}). However, most of these works discuss the causal relationship between the maps and racial residential divisions at the time, i.e. “How responsible are the HOLC maps for the residential segregation in the post-Depression era?” Few works directly address the period thereafter, what I will call the post-FHA era — the period after the signing of the Fair Housing Act of 1968, up until the present day — when the harms of redlining were recognized and legal and structural barriers were put in place to prevent discriminatory lending practices. The works which address this question of legacy (\citet{aaronson20}, \citet{aaronson21}, \citet{appel16}, \citet{YoungChacon2018}, \citet{faber20}, \citet{best22}) typically employ one of three techniques: (a) analyze various statistical outcomes (i.e. access to credit, recreation, etc.) across the risk zones and draw generalized conclusions about the role of the maps in producing those inequities, (b) analyze outcomes within a specified “buffer distance” of A-B, B-C, and C-D borders, under the idea that absent a HOLC map, variation across these buffer zones should be minimal, or (c) compare dissimilarity metrics (such as those introduced in \citet{MasseyDenton1988}) across appraised and unappraised cities. However, method (a) fails to distinguish between the fact and the method of drawing districts – that is, it tests against a null hypothesis of no variation between zones whereas a more powerful null hypothesis would be an alternative spatial variation. Method (b) covers the counterfactual by addressing the effect of the borders directly, but sacrifices some generalizability in the process – it requires the arbitrary selection of “buffer zones” with a fixed width around each of the boundaries in a map. Furthermore, this approach assumes but does not demonstrate the boundaries matter. Method (c) appears the closest to this method and addresses the question of causality. However, due to the widespread use of HOLC appraisals in major US cities, the unappraised counterfactual cities are limited. Crucially, prior literature does not consider the specific boundaries independent of the HOLC grade.

This paper provides two main contributions to the current discourse. First, I employ a Markov Chain Monte Carlo (MCMC) method of randomly generating plausible “counterfactual” maps representing possible alternative borders between risk areas, increasing confidence in prior findings. While methods (a) and (b) discussed above show that variance exists across boundaries in the HOLC maps, my findings are the first to show strong evidence that the HOLC neighborhoods had “hard” edges that correspond surgically well to social boundaries that still shape the patterns of residence in cities. Second, the MCMC method is highly reproducible and reusable with fewer parameters which require manual tuning (such as the thickness of the buffer zones) meaning it can be applied at scale to multiple cities. While the methodology of most other studies requires city-specific knowledge, MCMC is more generalizable. This study includes analysis of 11 US cities/boroughs (Atlanta, Baltimore, Brooklyn, Bronx, Chicago, Detroit, Manhattan, Oakland, Philadelphia, Queens, San Diego), exceeding the scope of other studies. Finally, the usage of Shannon’s entropy allows the deeper exploration of the boundaries themselves – while other studies analyze the grades and the within-neighborhood racial distributions, this study is singularly focused on the boundaries drawn in the HOLC maps themselves.

\section{Background}

“Redlining” refers to the “discrimination that bases credit decisions on the location of a property to the exclusion of characteristics of the borrower or property” (\citet{Hillier2003}). In these maps, created for most large American cities at the time, neighborhoods which were considered most desirable were colored in green on the map and given an ‘A’ grade. Second-grade neighborhoods were colored blue and given ‘B’, third-grade yellow and ‘C’, and the worst-quality neighborhoods colored red and given a ‘D’ grade. The creation of these districts involved human surveyors observing each city block, and considering factors such as rent percentage, ethnic minority population, and average income level before assigning a grade to a neighborhood (\citet{glasker03}, \citet{NelsonWinlingMarcianoConnolly2021}). At the time of their creation, the HOLC redlining maps were widely considered to plot out the ‘good’ and ‘bad’ parts of a city (\citet{best22}).

The grades assigned to neighborhoods were based on the predominant theory at the time of growth and decay in residential neighborhoods — at first, neighborhoods would exhibit a trend of positive growth, resulting in the influx of white, high-income professionals, but would later inevitably decline, as professionals moved elsewhere and the neighborhood became predominantly minority-owned and low-income (\citet{Hillier2005}). In the eyes of the FHA and the HOLC, lending risk was directly tied to this trend, as high-income professionals would be less likely to default on loans (\citet{Hillier2005}). There is historical precedent for such a trend, as a similar period of growth and “decline” befell Brooklyn, which began first as a residential community for Manhattan urbanites (\citet{Jackson1987}). Lenders at the time would have been aware of Brooklyn and similar examples across the nation when they formulated this theory. By 1940, the neighborhood risk maps had become quite prevalent – the HOLC had produced maps for over 150 cities across the United States (\citet{MitchellFranco2018}). 

After the Civil Rights Movement of the 1960s brought attention to the practice of redlining, it was explicitly outlawed in the Fair Housing Act of 1968, which stated:
\begin{quote}
    As made applicable by section 803 of this title and except as exempted by sections 803(b) and 807 of this title, it shall be unlawful--(a) To refuse to sell or rent after the making of a bona fide offer, or to refuse to negotiate for the sale or rental of, or otherwise make unavailable or deny, a dwelling to any person because of race, color, religion, sex, familial status, or national origin. (\citet{Johnson1968})
\end{quote}
Although this law made it illegal to officially practice redlining, the damage had already been done. Many of the same racial residential divides present in the HOLC maps remain prevalent today, manifesting themselves in access to recreational and healthcare facilities (\citet{YoungChacon2018}), credit access (\citet{appel16}), business loans (\citet{Immergluck2002}), and private investment (\citet{gotham00}).

\section{Methodology}
The task of finding a relationship between the borders drawn in the redlining maps and the borders of residential segregation in present-day cities must fundamentally answer the question: “How closely does segregation today match the specific boundaries drawn in the HOLC maps?” 

In order to quantitatively answer the above question, we must computationally generate a set of counterfactual boundaries against which to compare the original boundaries. In designing this methodology, we must acknowledge that many factors could have influenced the distribution of residents post-HOLC. Such factors include the 1968 Fair Housing Act, immigration after the 1965 Immigration and Nationality Act (\citet{OngLiu1994}), and urban renewal (\citet{avila09}). The chosen methodology must be robust to these events.

The best solution is to represent the map generation process as a discrete-time Markov Chain. A Markov Chain is a stochastic process in which the current step can be probabilistically computed from only the previous step, without any other prior information (\citet{bremaud20}). If we treat each map as a state in the chain, we can programmatically make small changes to the borders. In the long run, this will create a distribution of the ways borders could have been drawn. However, it is important to note that this is not a direct counterfactual of the maps – we are only choosing to analyze the borders, not the grades themselves.

I can summarize the methodology in five steps:

\begin{figure}
	\centering
	\includegraphics[width=\linewidth]{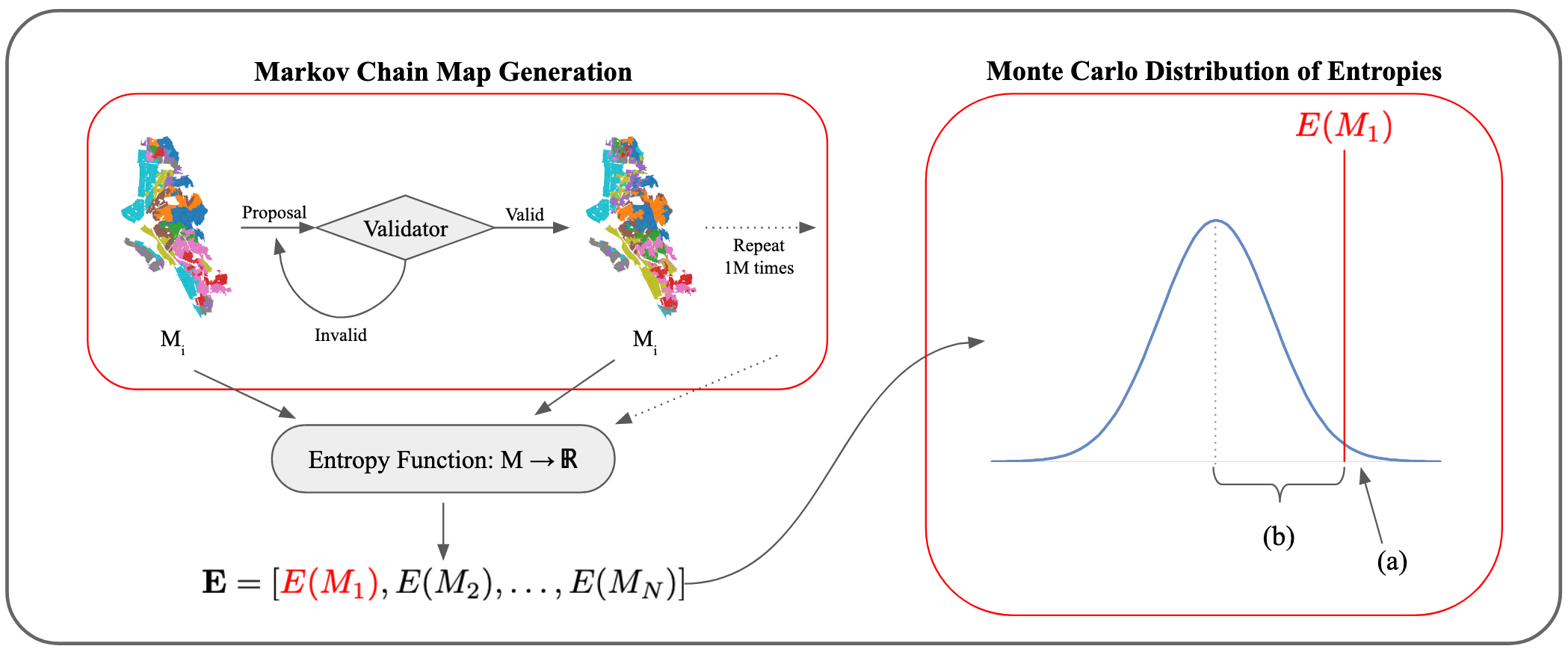}
	\caption{Graphical overview of methodology.}
	\label{fig:fig1}
\end{figure}

\begin{enumerate}
    \item Generate a large number of randomly generated maps with districts of equivalent size to the HOLC districts. Maps are only considered if they pass some predetermined validation criteria (as described in III.C), thus ensuring a baseline level of plausibility. These maps are generated until the statistic computed in step 3 satisfies a convergence criterion.
    \item For each fifth “counterfactual” map, map the current racial population distribution onto the borders drawn, and then compute a metric of segregation (detailed in section III.E).
    \item Using the measures calculated in step 2, assemble a univariate distribution of the “typical” levels of segregation reflected under the condition of borders for the given city. This will be different for each city due to geographical limitations that enforce certain boundaries (lakes, coastlines, mountain ranges, etc.).
    \item Steps 1-3 are repeated with additional Markov chains (with different random seeds) to confirm the distributions. The entropies generated from these additional chains are aggregated with the entropies from the first chain.
    \item Determine if the segregation score computed from the original HOLC maps is within or outside of the distribution computed in step 3.
\end{enumerate}

This process will result in a univariate distribution of some statistic along with a “test” value belonging to the initial HOLC maps. The null hypothesis is that the test value belongs to the distribution, and the alternative hypothesis is that it does not – indicating that the HOLC boundaries are descriptive of current racial residential divides. Figure \ref{fig:fig1} graphically describes the process.

\subsection{Markov Chain Monte-Carlo}

There exists a “cold start” problem with generating a library of alternative maps – we only have one sample map for each region being tested – the HOLC map. Thus, all alternative maps must somehow be “seeded” from this one in order to ensure they are plausible. Furthermore, it is computationally expensive to produce each new map from scratch – preliminary attempts failed to establish computational tractability. A Markov Chain solves the “cold start” problem with map generation – only a single map is needed to start the process, and all future maps are simply a variation of the map prior to it. In order to reduce dependence on the initial map, a “burn-in” period of 10\% of the total length of the Markov Chain is implemented, during which the chain is run but all outputs are discarded. A thinning factor of five is also used in order to improve computational efficiency and ensure better mixing of the chain. Using thinning, entropy statistics are only calculated on every fifth map in the chain, essentially allowing for four “burn in'' maps to be produced between every step. Overall, the 100 chains are run for ten thousand steps each, making for one million entropy measurements per region.

Overall, time and space efficiency are significantly superior to other methods. When the Markov Chain is run for a large number of steps, the distribution of the output can reveal information about the sample space (\citet{bremaud20}). This makes the chosen methodology a Markov Chain Monte Carlo process, and we can accurately (a) assume the entropy distribution is representative of all plausible border permutations and (b) traverse the sample space of all maps, provided the length of the Markov Chain is sufficiently large (\citet{MattinglyVaughn2014}). Similar methodologies have been designed to evaluate congressional district maps for instances of partisan gerrymandering (\citet{MattinglyVaughn2014}, \citet{fifeld20}, \citet{deford19}), but this is the first paper to extend the methodology to neighborhood risk maps.

\subsection{Case Selection}

Although the methodology presented in this section could feasibly be applied to any other city (provided data is available) in the United States, eleven cities/boroughs across the United States were selected for this study. The eleven regions selected were: Atlanta, Baltimore, Bronx, Brooklyn, Chicago, Detroit, Manhattan, Oakland, Philadelphia, Queens, and San Diego. All of these regions are confined to a single county, making data acquisition relatively straightforward, with the exception of Atlanta and Philadelphia, the former which required stitching data from Georgia’s Fulton, DeKalb, Milton, and Campbell counties\footnote{Milton and Campbell Counties were both annexed by Fulton County in 1932, thus they appear in the 1940 Census data but not in the 2020 dataset.} and the latter which incorporated Philadelphia and Montgomery counties.

These cities were selected to represent a wide range of cultural, geographic, and demographic histories, and to test if the methodology would provide similar results across various regions. Some of the cities, such as Philadelphia, Baltimore, and Detroit, were also selected for their significant recent and historical racial tensions as well as the wealth of literature available on the housing policies of the cities.

The demographic data for each of the cities was provided by the IPUMS project at the University of Minnesota in the form of 2020 block-group-level shapefiles (\citet{Manson2020}). Shapefiles of the HOLC boundaries were sourced from the Mapping Inequality project at the University of Richmond (\citet{NelsonWinlingMarcianoConnolly2021}).

\begin{figure}
	\centering
	\includegraphics[width=\linewidth]{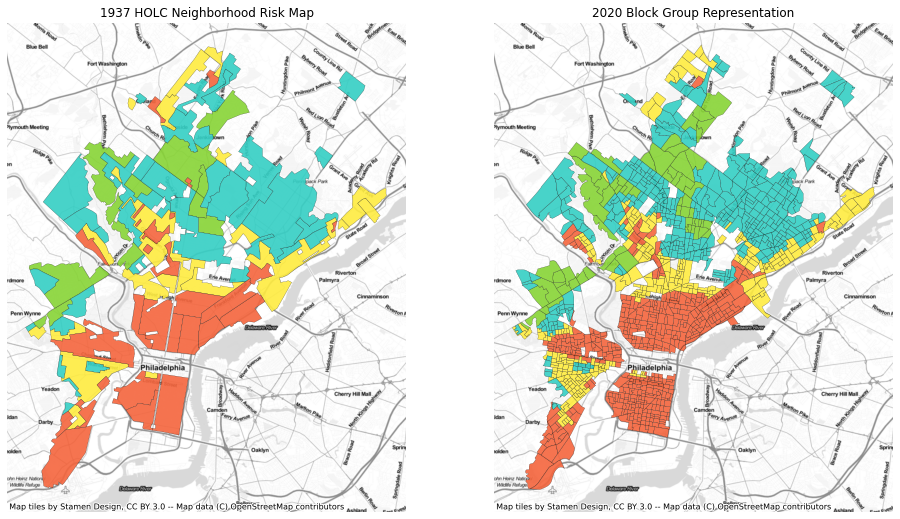}
	\caption{Crosswalk generation for Philadelphia, PA}
	\label{fig:fig2}
\end{figure}

Both demographic and city/region boundary data from 2020 were required for the map generation. Although the actual boundaries of the districts in the HOLC maps do not correspond to any US Census statistics subdivision boundaries, an approximation can be achieved using the US Census block group, as it is small yet statistically stable enough to avoid large amounts of uncertainty in data at the block level. No 1940 Census data was required, because the purpose of this study is to analyze the similarity between the HOLC maps and present-day racial residential boundaries, not the immediate short-term (5-10 years) effect of the maps. In order to first make the HOLC maps for each city compatible with a MCMC analysis, a “crosswalk” between the HOLC districts and the 2020 Census blocks was computed. A block was considered “valid” if 50\% of its land area was inside any HOLC district. Valid blocks are then assigned to the HOLC district with the greatest land area overlap. This produces a representation of the HOLC maps in terms of 2020 Census blocks that can then be fed into the Markov Chain for further analysis. Figure \ref{fig:fig2} depicts the crosswalk generation on the map of Philadelphia, Pennsylvania.

\subsection{Map Generators}

When the initial HOLC maps were drawn, they were not drawn using Census block group boundaries (Hillier 2005). This meant that, while the best available data for this study consisted of Census block groups for each region, the Markov Chain would need to generate maps that still approximated the boundaries of the HOLC districts. The proposal method used here is called chunk flip. Two neighboring districts are considered, and a random “chunk” of block groups is proposed to be flipped from one district to the neighboring one. This process repeats for a random number of the total district edges present in a given map. 

\subsection{Validators}

In addition to the map generator, the Markov Chain also employs a validator engine. The validator checks each generated map against a set of heuristics in order to ensure that maps do not diverge significantly from the plausible sample space. The specific validators are enumerated below:

Let $\bar{n}^{(t)}$ and $s^{(t)}$ be the mean and standard deviation of the district populations of the $t$-th map in the Markov Chain, where $t=0$ refers to the initial HOLC map.

\begin{enumerate}
    \item Lower Bound Control: $\min_{i = 1, \dots, z} \left\{ n_i^{(t)}\right\} \geq \min_{i = 1, \dots, z} \left\{ n_i^{(t)}\right\} mod~50$. This check ensures that there are no empty districts output in the map. Since the HOLC maps used to seed the algorithm are made by overlaying 2020 Census blocks on the 1940 districts, empty districts can arise when the population has shifted and abandoned regions that were previously districted in the original HOLC maps from the 1930s. Empty districts can complicate the entropy calculation done once the map has been validated, so all empty districts in each HOLC map are removed before the Markov Chain begins.
    \item Standard Deviation Control: $s^{(t)} \in \left[\frac34 s^{(0)}, \frac54 s^{(0)}\right]$. The purpose of this check is to verify that the populations of all districts in a map are somewhat similar. One assumption of the algorithm is that all districts within a city have more or less similar population counts, as an even distribution would make the maps most effective at achieving their primary goal of standardizing neighborhood-based risk maps. In addition, it is important to note that because the maps are drawn from the 2020 population distributions, which are unique from the 1940 population distributions, the chain will be more likely to divide on 2020 population patterns, creating an induced pressure towards the null hypothesis.
\end{enumerate}

A map must pass both of these conditions in order to be considered “valid” and for the Markov Chain to proceed. A validator that enforced shape constraints was considered and tested; however, it introduced significant computational overhead in having to verify each shape produced in each map from the Markov Chain. Instead, the minimum and average Polsby-Popper statistics were checked at the beginning and end of the chains to ensure that a significant amount of compactness was retained from the original maps and that the shapes had not diverged too irregularly. All eleven cities tested did not exhibit atypical shape behavior in the maps generated from the chains (see Appendix B).

\subsection{Entropy}

Once a map is deemed valid, the Monte Carlo section of the algorithm begins. There are multiple ideal measures of segregation I can compute from these maps, each measuring different dimensions of the notion of segregation: unevenness, exposure, concentration, centralization, and clustering (\citet{MasseyDenton1988}). The most common of these statistics would be the Index of Dissimilarity, measuring unevenness. The index of dissimilarity seeks to quantify the percentage of a minority population who would have to move to a new location to achieve an even distribution within the region. However, this index can only be used to describe the spatial unevenness of two racial groups, and thus fails to accurately describe the spatial unevenness of a whole region, populated by multiple racial groups. Instead, for each map, a region-wide “entropy” index, also known as a Shannon index (\citet{MasseyDenton1988}), is computed. This statistic will measure the unevenness of the racial distribution in a region:

Let $k$ be the number of racial groups. The 2010 Census has seven demographic options listed: White, Black, Asian, American Indian/Alaska Native, Hawaiian/Pacific Islander, Other, and “Two or more” groups. Since the Hispanic category is not given as an option in the Census race question but rather as its own question, I include the Hispanic population as a count of those who responded “yes” to the Census question. Thus, $k=8$. $p_{ij}$ will refer to the proportion of the $j$th racial group in the $i$th district ($p_{ij}=n_{ij}/n_i$).

First, $h_i$, the district-wise entropy index for district $i$, is calculated as follows:
\[h_i = -\sum_{j = 1}^k p_{ij}\ln{(p_{ij})}\]
Where $\ln(0) = 0$. With all district-wise entropies calculated, the region-wise entropy $H$ is calculated from $\hat{H}$, the region-wide entropy if the whole region is considered a single district, and $\bar{H}$, the population-weighted average of all $h_i$:
\[\hat{H} = -\sum_{j = 1}^k \left[
    \left(
        \frac{1}{\sum n_i} \sum_{i=1}^z n_{ij} 
    \right)
    \ln\left(
        \frac{1}{\sum n_i} \sum_{i=1}^z n_{ij} 
    \right)
\right]\]

\[\bar{H} = \frac{1}{z} \sum_{i = 1}^z \frac{n_i}{\sum n_i} h_i\]

\[H = \frac{\hat{H} - \bar{H}}{\hat{H}}\]

Intuitively, $H$ becomes bounded by 0 and 1 (inclusive), where a score of 0 corresponds to the most racially homogenous region (each district has the same racial distribution as the whole region). A score of 1 is the opposite, where each racial group appears to be siloed into their own districts. The entropy functions as a sort of correlation metric between the borders and the racial distribution of a city – if entropy is high, then the borders do a good job of segregating the population along the lines of the chosen metric. 

\subsection{Code Availability}
Appendix A contains code used in this study. The simulations are all written in Python 3 (\citet{VanRossumDrake1995}) and employ a few key libraries: gerrychain\footnote{\href{https://github.com/mggg/GerryChain}{https://github.com/mggg/GerryChain}}  from the Metric Geometry and Gerrymandering Group, geopandas (\citet{Jordahl2020}), and the SciPy stack (\citet{McKinney2010}).

\section{Results}

\begin{table}
	\caption{Results of all chains.}
	\centering
	\begin{tabular}{r|l|l|l|l|l|}
		\toprule
		City     & Baseline     & Mean (all chains) & Mean Absolute Difference (AD) & t value & $Pr\geq|t|$ \\
		\midrule
		Atlanta & 0.190 & 0.178 & 0.011 & -673.0 & 0.000* \\
        Baltimore & 0.201 & 0.143 & 0.059 & -417.9 & 0.000 \\
        Brooklyn & 0.222 & 0.110 & 0.027 & -534.2 & 0.000 \\
        Bronx & 0.137 & 0.182 & 0.039 & -426.9 & 0.000 \\
        Chicago & 0.371 & 0.350 & 0.021 & -776.1 & 0.000 \\
        Detroit & 0.417 & 0.368 & 0.049 & -740.3 & 0.000 \\
        Manhattan & 0.187 & 0.167 & 0.021 & -411.9 & 0.000 \\
        Oakland & 0.130 & 0.120 & 0.009 & -664.8 & 0.000 \\
        Philadelphia & 0.194 & 0.155 & 0.033 & -372.4 & 0.000 \\
        Queens & 0.250 & 0.237 & 0.013 & -580.1 & 0.000 \\
        San Diego & 0.163 & 0.147 & 0.017 & -351.0 & 0.000
	\end{tabular} \\
    $*:~< 2.2 \times 10^{-16}$
	\label{tab:table}
\end{table}

\begin{figure}
	\centering
	\includegraphics[width=\linewidth]{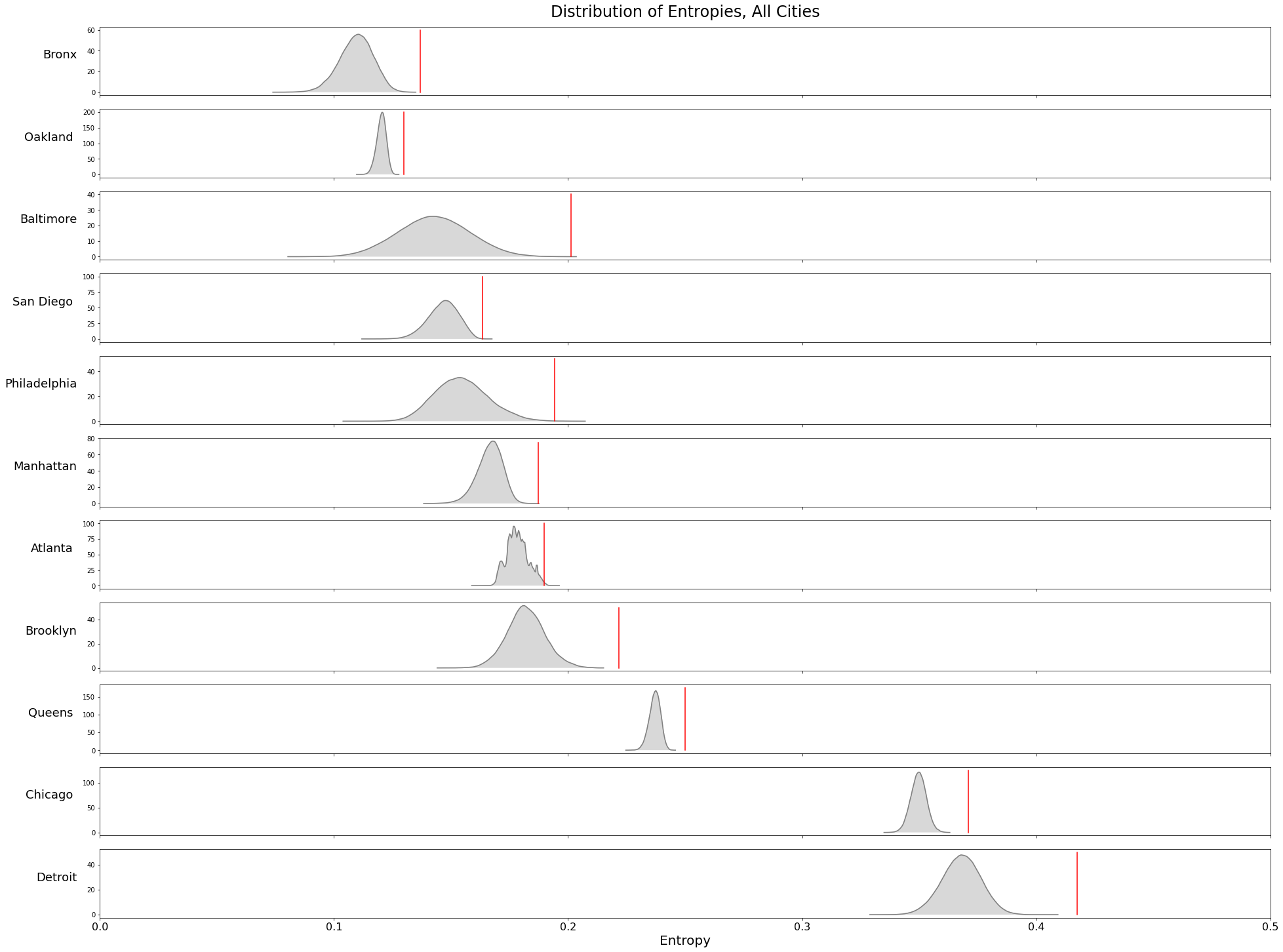}
	\caption{Distribution of entropies, all cities}
	\label{fig:fig3}
\end{figure}

Figure \ref{fig:fig3} depicts the univariate distribution of entropies from all 100 chains compared against the baseline entropy from the HOLC maps. As seen in Figure \ref{fig:fig3} and Table \ref{tab:table}, all of the 11 cities tested showed a clear difference between the HOLC map entropy and the entropies of the randomly generated maps. 

Convergence is an important aspect of Markov Chain Monte Carlo methods, and preliminary visualization indicates that Markov Chains in most regions did converge. Appendix C contains details about the statistical testing for convergence. However, in the absence of convergence across chains in a region, we can analyze the reduction of variance in individual Markov Chains. In this particular use case, there may exist multiple “equilibrium distributions” that the Markov Chains may find, such that not all of the chains will converge to the same entropy levels. By running 100 independent chains, we can maximize our chance of finding as many of these equilibria as possible. 

With the exception of Atlanta (which has an exceptionally sparse HOLC map relative to the modern city layout), the distribution of entropies across the 100 chains is relatively Gaussian. Therefore, I use a two-sided, one-sample t-test to measure the statistical significance of the baseline from the distributions. Table \ref{tab:table} shows the mean t-score and p-values for the t-tests run on each chain in each region. Overall, all cities produce statistically significant results.

Under the assumption that the entropies of all randomly generated maps in a region fall along approximately the generated distribution, the probability that the HOLC redlining maps describe a level of segregation equal to that of a randomly generated map is close to zero for all cities tested. Thus, there appears to be little to no evidence to support the null hypothesis that, across these cities, the 1930s HOLC redlining maps describe similar levels of racial residential segregation as any randomly generated, plausible map. The mean entropy from the Markov Chains is significantly lower than the HOLC entropies in all cities tested, which demonstrates that the HOLC maps better characterize the racial residential boundaries in their respective cities than any plausibly generated, random map as produced by the Markov Chain.

By analyzing the distance between the mean of the distribution and the HOLC maps’ entropy, we can also learn about the relative power of this test on each city. In the case of Detroit, for example, the absolute difference from the Markov Chain was 0.049, which is about 188\% higher than the absolute difference for San Diego (0.017). We know that the absolute difference is lower-bounded at 0, which makes it a useful indicator to compare the results of multiple cities. 

\subsection{Case Study: Philadelphia}

In order to provide better context around the values presented in the above section, we can investigate the case of Philadelphia, a city with a well-studied housing discrimination past. From 1935 to 1937, Philadelphia’s neighborhood risk map underwent several revisions before the final versions were published (\citet{Hillier2005}). The maps (see Appendix D) permuted the borders considerably – significant changes are visible in eastern and northern Philadelphia between the second and third iterations of the map, for example. These permutations were likely arbitrary and reflective of the moods of individual appraisers, as the grades follow the pattern of getting harsher despite Philadelphia’s improving economic climate in the 1930s, and there was significant personnel change between map versions (\citet{Hillier2005}). Moreover, the boundaries were not based on any existing setup (like the Census tract boundaries) (\citet{Hillier2005}), lending more credence to the fact that they were designed arbitrarily, around homogeneous neighborhoods and demographics.

When redlining was officially outlawed, as it was in other cities, in 1968 with the passage of the Fair Housing Act, Philadelphia underwent more radical changes to their residential demography than most other cities. The Community Reinvestment Act of 1977, which was designed to stymie the effects of redlining, allowed young white homeowners to move into “D” graded neighborhoods in recent years (\citet{best22}). After 1990, immigration also affected the city, resulting in the creation of ethnic neighborhoods such as Little Saigon in southern Philadelphia, where Vietnamese immigrants settled and were able to find support networks (\citet{eichel18}). Crucially, these changes still happened along the lines defined in the HOLC maps. For example, Little Saigon appears to fall squarely in the “D” graded D-17 district. State Highway 1 in northwestern Philadelphia also remains a common border with predominantly White and Black communities inhabiting the northern and southern regions divided by the freeway, respectively.

The results across the 100 chains run for Philadelphia are shown in the figure below. Over the length of the chain, the entropy statistic continues to decrease and eventually converges to 0.161, over three points lower than Philadelphia’s baseline, 0.194.

\begin{figure}
	\centering
	\includegraphics[width=\linewidth]{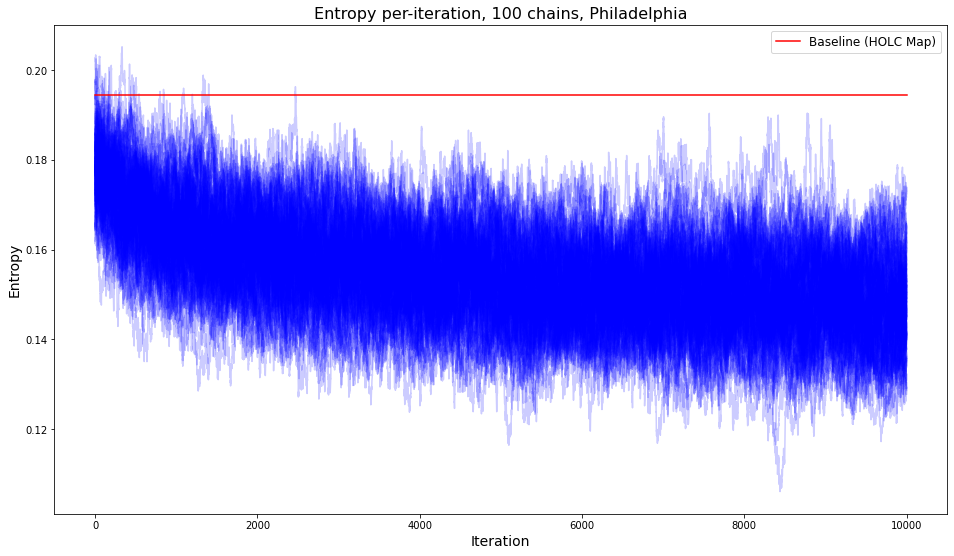}
	\caption{Entropy as a function of chain iterations for Philadelphia}
	\label{fig:fig4}
\end{figure}

In order to ensure that the chains’ convergence away from the baseline is not due to the methodology but the data itself, I constructed an ‘artificial dataset’ from the 2020 Census block group data – within each racial group, I shuffled the number of individuals present in any given block group. Doing this allowed the overall racial breakdown of the city to stay the same, while changing the distribution within the block groups. After seeding a new 100 Markov Chains with this ‘fake’ data, we can see that the entropy no longer diverges from the baseline. 

\begin{figure}
	\centering
	\includegraphics[width=\linewidth]{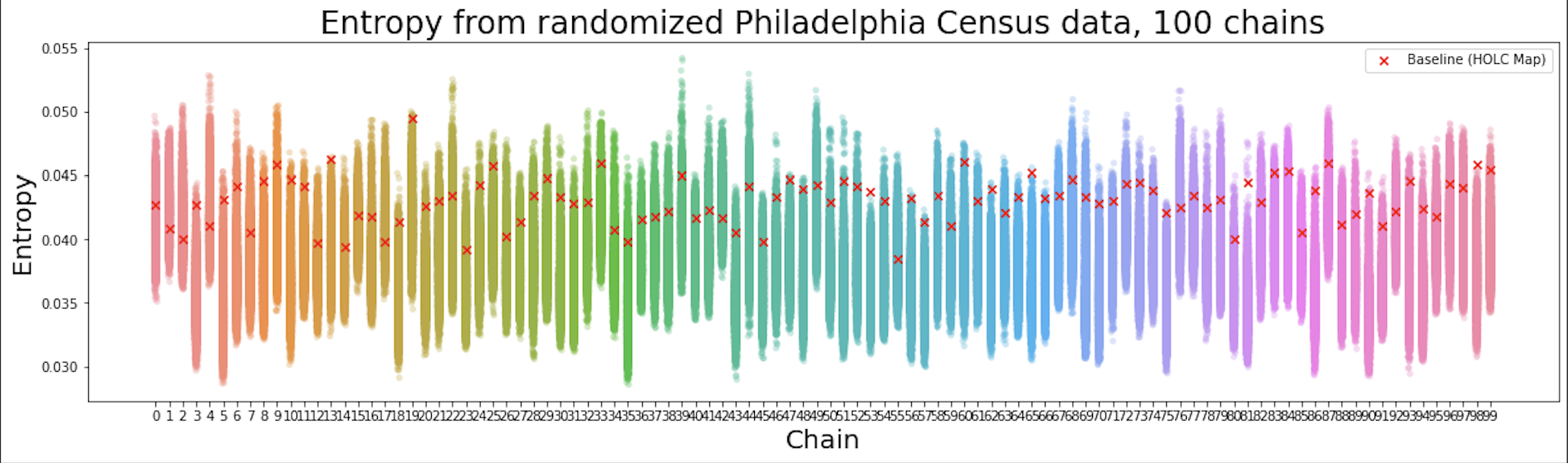}
	\caption{Distributions for all randomized chains for Philadelphia and their starting entropies (marked by a red 'x')}
	\label{fig:fig5}
\end{figure}

Unlike the original data, each of the 100 chains for the “fake” runs has its own baseline – we can see that while there are some outliers, the baselines generally fall within or towards the top of the entropy distribution. This analysis serves as a placebo test against the results in Figure \ref{fig:fig4}. There is a risk that the map generator and validators could bias the Markov Chains towards certain types of maps, wherein the generator will get “stuck” in a certain space of map proposals and thus return consistently low entropy scores. However, randomizing the data and showing that such risk is minimal indicates that such bias is not a property of the research design but rather the racial distribution of the city – and we have a greater assurance that the Markov Chains have traversed the space of possible maps well.

\section{Conclusion}

The results depicted in Section IV strongly indicate that the borders of the HOLC redlining maps persist, with surgical precision, into the present day. There are two potential interpretations from these results: first, that HOLC created the boundaries which persist to this day, and second, that HOLC managed to describe existing racial boundaries in 1930 with a suspiciously high level of precision. From either conclusion, however, the findings corroborate previous claims that HOLC borders effectively institutionalized lines of segregation, allowing them to carry momentum into the present day (\citet{dreier14}, \citet{HirschConnolly2021}, \citet{Jackson1987}, \citet{Hillier2003}, \citet{KrysanCrowder2017}, \citet{faber20}). In exclusively analyzing the borders presented in these maps and not the grades themselves, I am able to account for a variety of confounding factors. These include chain migration effects (\citet{flippen21}) and an increase in the immigration of non-Black racial minorities (such as Asians and Hispanics) into the US in the 1960s. Despite, for example, the high levels of immigration from East Asia in the late 1900s, we find that the borders presented in the HOLC maps still affected the way in which those populations settled. Additionally, despite integrationist housing policies such as The Fair Housing Act of 1968, The Equal Credit Opportunity Act of 1974, The Home Mortgage Discrimination Act of 1975 (\citet{Hillier2003}), and the Obama-era Affirmatively Furthering Fair Housing rules, the borders still persist, indicating that the intended effect of these policies was not necessarily achieved.

In showing that the specific boundaries from the HOLC maps are descriptive of current racial residential divides as opposed to racial divides between generic neighborhoods, this study builds on existing literature. Such a significant descriptive effect inherently rules out many counterfactuals such as other racialized housing programs or the effects of education or tax policies, to name a few examples.

This study also sets the stage for future MCMC-driven investigations into the persistence of redlining. Future researchers may consider testing different outcome statistics, such as educational attainment, welfare program dependence, or household income, to see if they also happen to correlate to the HOLC boundaries. Should readers wish to extend the findings of this study to specific racial minorities, it is suggested that one replaces the measure of entropy with the index of dissimilarity (\citet{MasseyDenton1988}) and then pick two racial groups to compare (i.e. White and Black Americans). The index of dissimilarity will allow for direct comparison between two racial groups, and can thus result in more granular and specific results than the one provided in this study.

Optimizations to the generator process exist, such as the recombination method proposed by \citet{deford19}, or the sequential Monte Carlo method designed by \citet{McCartanImai2020}. In the setting of legislative redistricting, these methods are more efficient at sampling from the target distribution of maps (\citet{fifeld20}). However, while recombination works well on contiguous maps such as voting precincts in states, it is ill-suited to maps with non-contiguous regions (such as islands) and would require investigator judgment to fix, removing one of the key benefits of the methodology.

Indeed, the rapidly reproducible nature of the methodology means it can be applied at different times, with different datasets (i.e. older Census data from 2000 or 1990) in order to test the efficacy of equal housing legislation in a specific region. This could make the methodology a useful “sanity check” for local governments to evaluate their progress in erasing the discriminatory effects of the HOLC redlining maps.

Combined, these insights are valuable contributions to the literature on the legacy of redlining. Although it does not indicate directly that the redlining maps were the source of the racial residential divides in American cities, it finds that in the post-FHA period the United States has effectively “colored inside the lines” – despite the remarkable amount of urban change since the maps’ publishing, American cities remain divided along the exact lines from the original maps. Such a finding is a useful reminder that little progress has truly been made to reverse the effects of the FHA, and that much work remains to be done to desegregate American cities (\citet{faber20}). 


\bibliographystyle{unsrtnat}
\bibliography{references}  






\appendix

\section{Appendix A: Code and Reproducibility}
All code for this project, including a complete reproducibility guide, can be found at \href{https://tinyurl.com/redliningcode}{https://tinyurl.com/redliningcode}.

\section{Appendix B: Shape Constraints}

A shape constraint based on the Polsby-Popper compactness metric was tested as a potential validator. The Polsby-Popper statistic measures the level of similarity between a polygon $D$ and a circle of the same area as $D$. I will denote the area of $D$ as $A(D)$ and the perimeter as $P(D)$.

Polsby-Popper is computed as follows:

\[PP(D)= \frac{4\pi A(D)}{P(D)^2}\]

In order to validate, the minimum Polsby-Popper score from the HOLC neighborhoods was computed as used as a baseline. This was deemed to be the minimum level of compactness acceptable for the region. At the end of each Markov chain, the minimum Polsby-Popper was computed again, and checked to see if it was lower than the baseline. Of the 1,100 Markov chains run (11 cities, 100 chains each), all chains passed this validation, indicating that the generator was not producing boundaries which were more irregular than those present in the HOLC maps.

\section{Appendix C: Statistical Testing for Markov Chain Convergence}
In order to empirically verify the convergence of the chains, I checked the Gelman-Rubin R-hat for these chains. The Gelman-Rubin statistic serves as a litmus test to measure the convergence of a Markov Chain by comparing the ratio of the across-chain and between-chain variances (\citet{Roy2019}). The formula can be seen below. When the Gelman-Rubin score is below 1.2, the chains are considered to have converged (\citet{brooks98}).

In order to calculate the Gelman-Rubin R-hat for a given region, we must compute a ratio of two terms, $W$ and $\hat{V}$. Let X be a m by n matrix where $X_{ij}$ is the $j$th observation (after burn-in) from the $i$th Markov chain.

\[E = \frac{\sum_{i=1}^m \sum_{j = 0}^{n-1} (X_{ij} - \bar{X}_{i\cdot})^2}{m(n-1)}\]

\[\hat{V} = \frac{W(n-1)}{n} + \frac{\sum_{i = 1}^m (\bar{X}_{i\cdot} - \bar{X}_{\cdot \cdot})^2}{m - 1}\]

\[R = \frac{\hat{V}}{W}\] 

The intuition is that W estimates the within-chain variance, and V-hat estimates between-chain variance. If between-chain variance is only marginally greater than within-chain variance (R-hat is 1.2), then the chains have converged to the same sample space. The table below reports the Gelman-Rubin R-hat values for each of the tested cities. Note, however, that the Gelman-Rubin score does not paint a complete picture of convergence – some cities may exhibit multiple equilibria. When within-chain variance is low and the R-hat is above 1.2, it would indicate that each Markov chain is independently finding its own “natural state” map which produces a certain entropy. At this point if we treat each of the 100 Markov chains as their own independent draws, then we can test the (100) converged entropy values against the baseline. Another option would be to run the chains for more than 10 thousand iterations, as the Gelman-Rubin R-hat appears to monotonically decrease as the iterations increase.

\begin{table}
	\centering
	\begin{tabular}{r|l}
		\toprule
		City     & R \\
		\midrule
        Atlanta & 2.676 \\
        Baltimore & 1.175 \\
        Bronx & 1.156 \\
        Brooklyn & 1.146 \\
        Chicago & 1.436 \\
        Detroit & 1.515 \\
        Manhattan & 1.108 \\
        Oakland & 1.953 \\
        Philadelphia & 1.108 \\
        Queens & 1.151 \\
        San Diego & 1.688
	\end{tabular}
	\label{tab:aptable}
\end{table}

\section{Appendix D: Development of HOLC maps for Philadelphia, PA}
Source: \citet{Hillier2005}

\begin{figure}
	\centering
	\includegraphics[width=\linewidth]{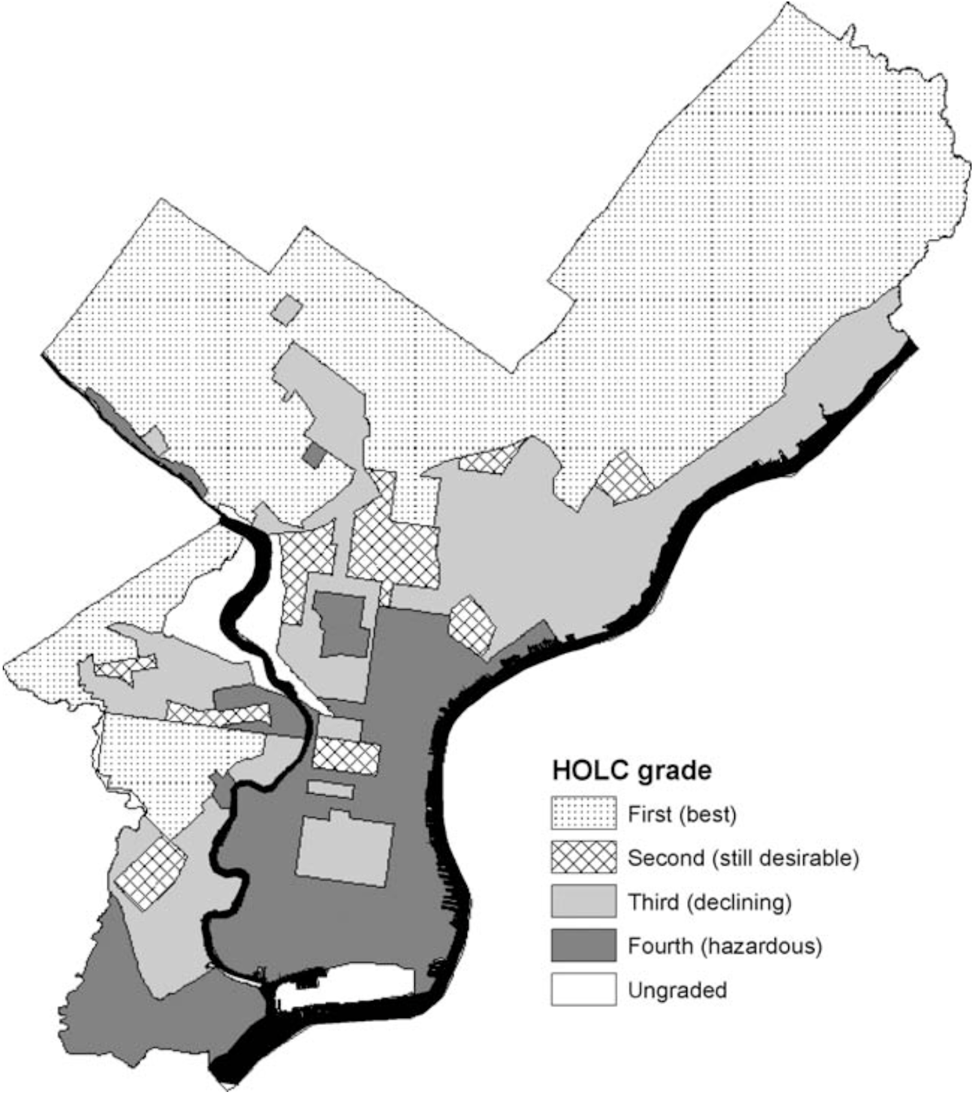}
	\caption{First HOLC neighborhood risk map, Philadelphia 1935}
	\label{fig:a1}
\end{figure}

\begin{figure}
	\centering
	\includegraphics[width=\linewidth]{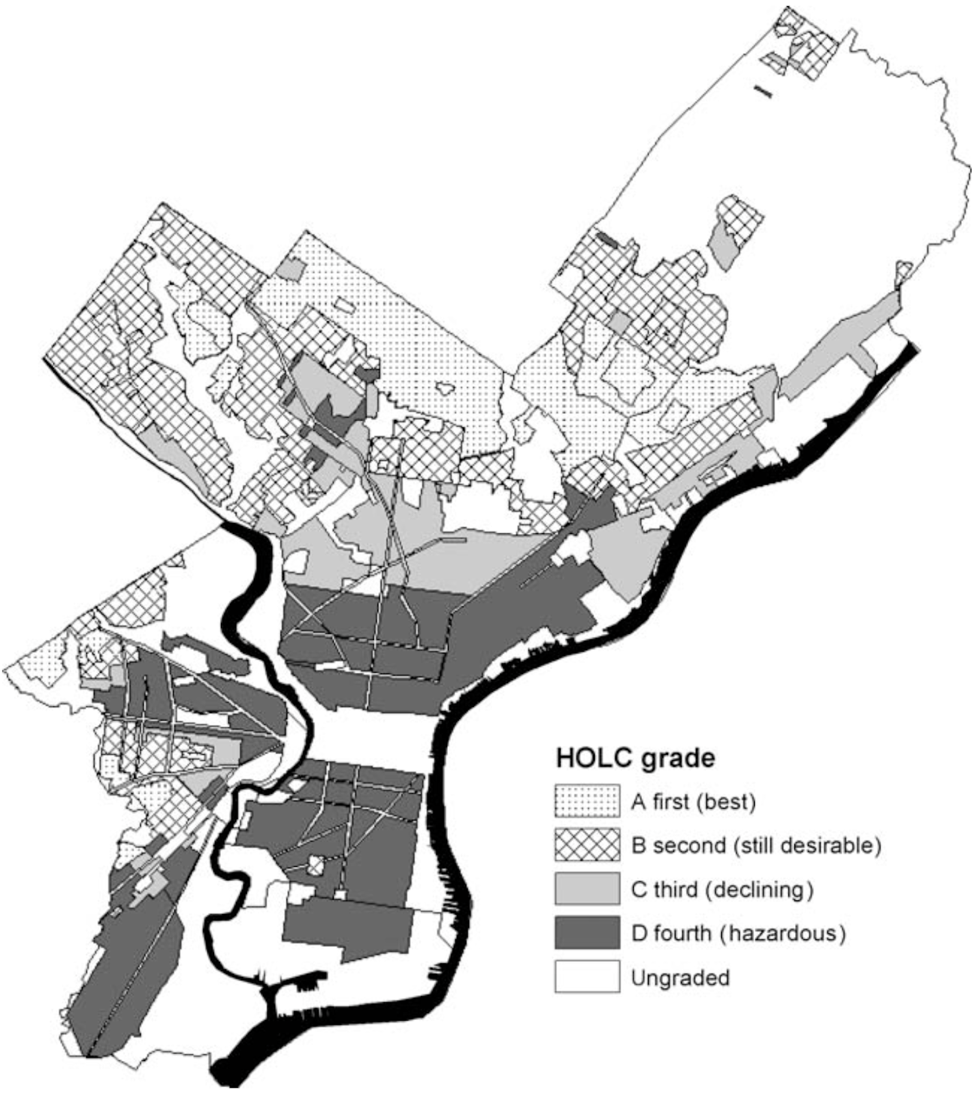}
	\caption{Second HOLC neighborhood security map, Philadelphia 1936}
	\label{fig:a2}
\end{figure}

\begin{figure}
	\centering
	\includegraphics[width=\linewidth]{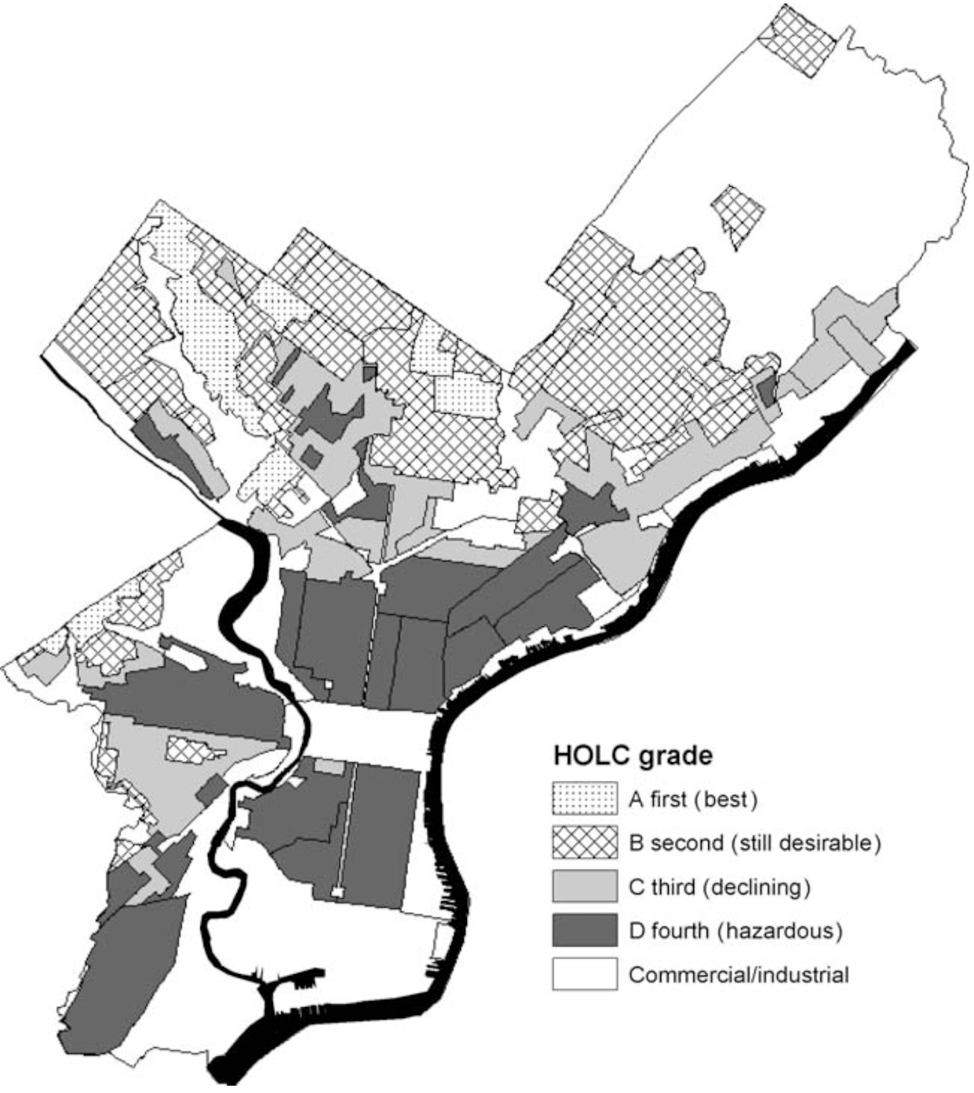}
	\caption{Third and final HOLC neighborhood security map, Philadelphia 1937}
	\label{fig:a3}
\end{figure}

\end{document}